\documentclass[12pt,onecolumn,letter]{article}
\usepackage{graphicx}
\usepackage{authblk}
\usepackage[margin=0.7in,letterpaper]{geometry}
\usepackage{subcaption} 

\title{Compton-Pair Production Space Telescope (ComPair) for MeV Gamma-ray Astronomy}

\author[1,*]{A.A. Moiseev}
\author[2]{M. Ajello}
\author[3]{J.H. Buckley}
\author[4]{R. Caputo}
\author[1]{E.C. Ferrara}
\author[2]{D.H. Hartmann}
\author[5]{E. Hays}
\author[5]{J.E. McEnery}
\author[5]{J.W. Mitchell}
\author[1]{R. Ojha}
\author[5]{J.S. Perkins}
\author[5]{J.L. Racusin}
\author[1]{A.W. Smith}
\author[5]{D.J. Thompson}

\affil[1]{CRESST/NASA/GSFC and University of Maryland, College Park, Maryland 20771, USA}
\affil[2]{Department of Physics \& Astronomy, Clemson University, SC 29634, USA}
\affil[3]{Department of Physics, Washington University, St. Louis, MO  63130, USA}
\affil[4]{UCSC and SCIPP, Santa Cruz, CA 95064, USA}
\affil[5]{NASA/GSFC, Greenbelt, Maryland 20771, USA}
\affil[*]{e-mail: alexander.a.moiseev@nasa.gov}

\date{}

\begin{document}
\maketitle       

\abstract{The gamma-ray energy range from a few hundred keV to a few hundred MeV has remained largely unexplored, mainly due to the challenging nature of the measurements, since the pioneering, but limited, observations by COMPTEL on the Compton Gamma-Ray Observatory (1991-2000). This energy range is a transition region between thermal and nonthermal processes, and accurate measurements are critical for answering a broad range of astrophysical questions. We are developing a MIDEX-scale wide-aperture discovery mission, ComPair (Compton-Pair Production Space Telescope), to investigate the energy range from 200 keV to $> 500$ MeV with high energy and angular resolution and with sensitivity approaching a factor of 20-50 better than COMPTEL. This instrument will be equally capable to detect both Compton-scattering events at lower energy and pair-production events at higher energy. ComPair will build on the heritage of successful space missions including Fermi LAT, AGILE, AMS and PAMELA, and will utilize well-developed space-qualified detector technologies including Si-strip and CdZnTe-strip detectors, heavy inorganic scintillators, and plastic scintillators. }

\normalsize

\section{Introduction }

The MeV domain is one of the most underexplored windows on the Universe. This has not resulted from a paucity of interesting science, but from technology 
constraints that have limited advances detection sensitivity in recent decades. Thanks to currently available technology that makes sensitive observations in this band feasible, we can address this tantalizing gap in our knowledge pertaining to critical astrophysical questions. Known science that can be addressed at these energies is much too broad for any realizable single mission. Hence we plan to focus on a large field of view instrument with moderate angular and energy resolution optimized for continuum sensitivity. We have chosen to optimize our instrument  in this energy range, one we call ComPair, to use both Compton scattering and pair production as detection techniques, 
for a compelling, high scientific return design. Such design can be achieved at reasonable cost with technologies that are currently in a high state of readiness.

Measurement at these energies is challenging. This is mainly due to specifics of photon detection: it is a range where two processes of photon interaction, Compton scattering and pair production, compete, with a crossover at around 10 MeV depending on the material. These two interaction processes require different approaches in both detection and data analysis, and consequently in the instrument concept. The ideal would be to have two separate instruments, each optimized for a particular photon interaction. However, we believe that it is possible, though challenging, to design a cost-saving single instrument that will be capable to detect both kinds of photon interaction processes and provide accurate results in the extended energy range 0.2  -  500 MeV, optimizing its performance in the 0.5 - 100 MeV span. An additional advantage of such an approach is that the Compton and pair-production events will be treated separately, and the results will be compared in the overlapping energy range. This will significantly increase the confidence in the results obtained, and improve the assessment of systematic errors.

\section{ComPair Instrument Concept}
 The instrument concept is based on the traditional scheme of previous gamma-ray telescopes (e.g. EGRET \cite{egret}, Fermi-LAT  \cite{glast}, AGILE  \cite{agile}), as well as on the engineering studies for the European-led MEGA \cite{mega} and GRIPS  \cite{grips} projects  and utilizes well-known detectors without any consumables \cite{compair}.  It consists of a multi-layer silicon-strip Tracker in which the gamma-rays interact and the resulting charged particles have directions and energy measured; a Calorimeter to measure deposited energy and position of photon absorption, consisting of a novel combination of solid-state detectors CdZnTe-strip (CZT), and inorganic scintillators CsI(Tl); and a surrounding plastic scintillator anticoincidence detector (ACD), to discriminate the photons from the vastly more numerous charged particles in the space environment (Fig. 1).

\begin{figure}[ht]
\centering
\includegraphics[width=0.7\textwidth]{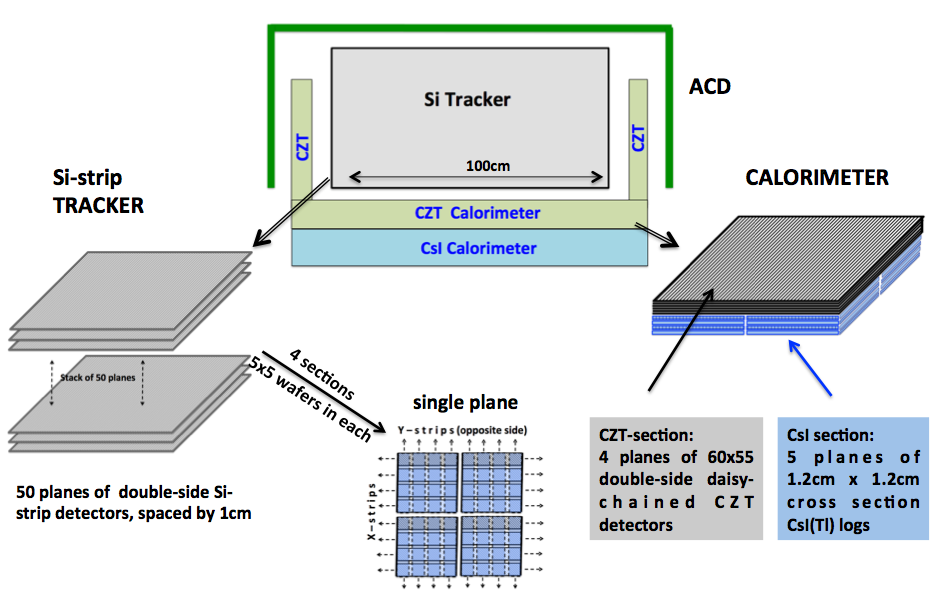}
\caption{Conceptual design of ComPair}
\end{figure}

\begin{figure}
\centering
\includegraphics[width=0.6\textwidth]{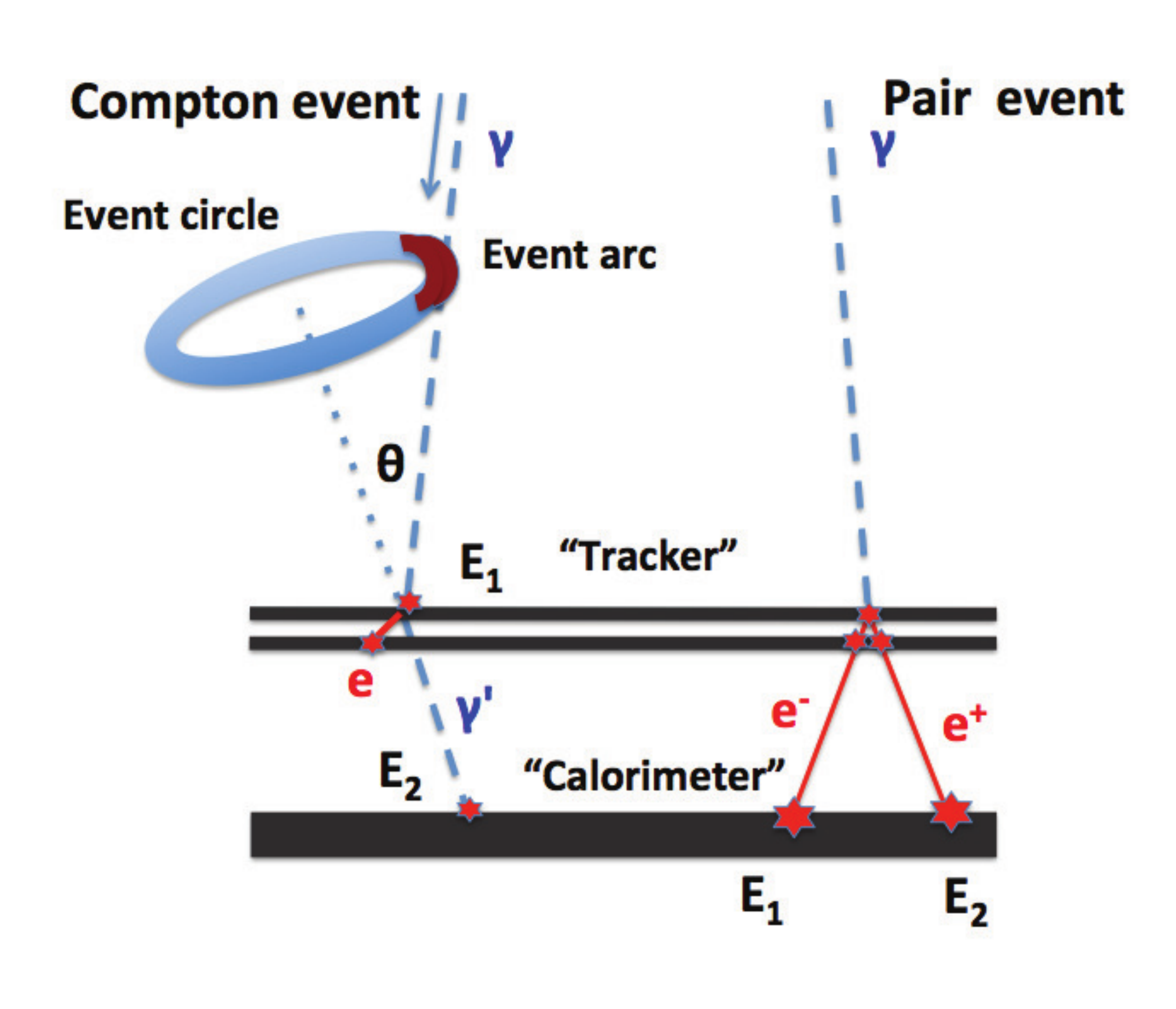}
\caption{Event types for Compton-Pair telescopes.}
\end{figure}

For pair-production events, ComPair is similar in design to AGILE and Fermi-LAT, but optimized for lower energy. This goal can be achieved by eliminating the passive tungsten converters used in both these instruments. This approach reduces gamma-ray conversion efficiency (roughly proportional to the tracker layer thickness), but it improves the instrument point-spread function (PSF) by reducing absorption and multiple Coulomb scattering of the electron and positron.
The broad PSF is a primary limiting factor in the science that can be done at energies below 100 MeV by AGILE and Fermi-LAT. In addition to improving the PSF, the use of low-mass tracker planes also enables photon polarization measurements.

Detecting gamma-rays by Compton scattering (the dominant process below about 10 MeV) is more complicated than using pair production, because the scattered photon carries a significant amount of the information about the incident photon and so needs to be detected too. In practice, a Compton telescope requires two separate photon interactions in order to have a clear detection. In the ComPair concept, the first Compton scattering of the incident photon occurs in one of the Tracker planes, creating an electron and a scattered photon. The Tracker measures the interaction location, the electron energy, and in some cases the electron direction. The scattered photon can be absorbed in the Calorimeter or scattered a second time in the Tracker before being absorbed in the Calorimeter where its energy and absorption position are measured.

The basic principle of the Compton mode of operation is illustrated in Figure 2, left. An incident gamma-ray Compton scatters by an angle $\Theta$ in one layer of the Tracker, transferring energy $E_{1}$ to an electron. The scattered photon is then absorbed in the Calorimeter, depositing energy $E_{2}$, and 
the scattering angle is given by $\cos\Theta = \frac{m_{e}c^{2}}{E_2} + \frac{m_{e}c^{2}}{E_1+E_2} $, where $m_ec^2$ is the rest mass energy of the electron. 
With this information, we derive an "event circle" from which the original photon arrived. We will call this sort of Compton events "untracked" events. The uncertainty in the "event circle" reconstruction is reflected in its width and is due to the uncertainties in direction reconstruction of the scattered photon and the energy measurements of the scattered electron $(E_1)$ and the scattered photon $(E_2)$. If the scattered electron direction is measured, the event circle reduces to an event arc with length due to the uncertainty in the electron direction reconstruction, allowing improved source localization. This event is called a "tracked" event, and its direction reconstruction is somewhat similar to that for pair event: the primary photon direction is reconstructed from the direction and energy of two secondary particles: scattered electron and photon.  

\subsection{Si-strip Tracker}
Silicon strips have become the standard particle tracking technology on the ground and in space.  Double-sided silicon detectors (DSSD) are necessary for measuring both x- and y-coordinates of Compton event starting points (low-energy Compton electrons may not penetrate two single-sided detectors) and for measuring polarization for pair-production events: 

\begin{itemize}
\item Analog readout is necessary for measurement of the energy of the scattered Compton electron, as well as for measuring energies of pair-production events at low energy. 
\item The thickness of Si needs to be optimized between conversion efficiency (affects instrument effective area) and multiple scattering (affects PSF). 
\item Si-strip pitch needs to be optimized between PSF (for Compton events, and for pair-production events at higher energy) and number of electronics channels (cost, power). 
\item Si-strip plane spacing needs to be optimized between PSF for pair-production events at higher energy and the instrument field of view. 
\end{itemize}

The Tracker is a stack of 50 double-sided 0.5 mm thick Si-strip detector planes, with an area of 1m x 1m and strip pitch 0.25mm. The planes are separated by 1 cm, which we currently consider optimal in terms of the PSF/field-of-view trade-off. Each plane is made of 10 x 10 DSSD wafers, 9.5cm x 9.5cm each, divided in four 5 x 5 wafer segments with daisy-chained strips in each direction (within one segment). In this layout the readout is arranged on the Tracker sides (tentatively by IdEAS VATA4560.3), without the need to have mechanically separated "towers" (Fig. 3).

\begin{figure}[ht]
\hspace{1cm}
\begin{minipage}[t]{0.4\linewidth}
\includegraphics[width=\textwidth]{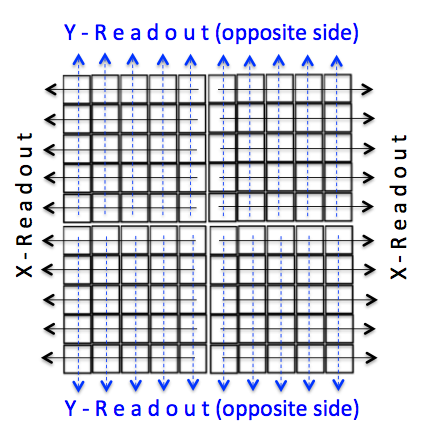}
\caption{Si Tracker single plane, divided in 4 segments, with 5 x 5 wafers in each (25 x 25 for CZT plane)}
\label{fig:figure1}
\end{minipage}
\hspace{2mm}
\begin{minipage}[t]{0.45\linewidth}
\includegraphics[width=\textwidth,trim=0 0 30 0 ,clip=true]{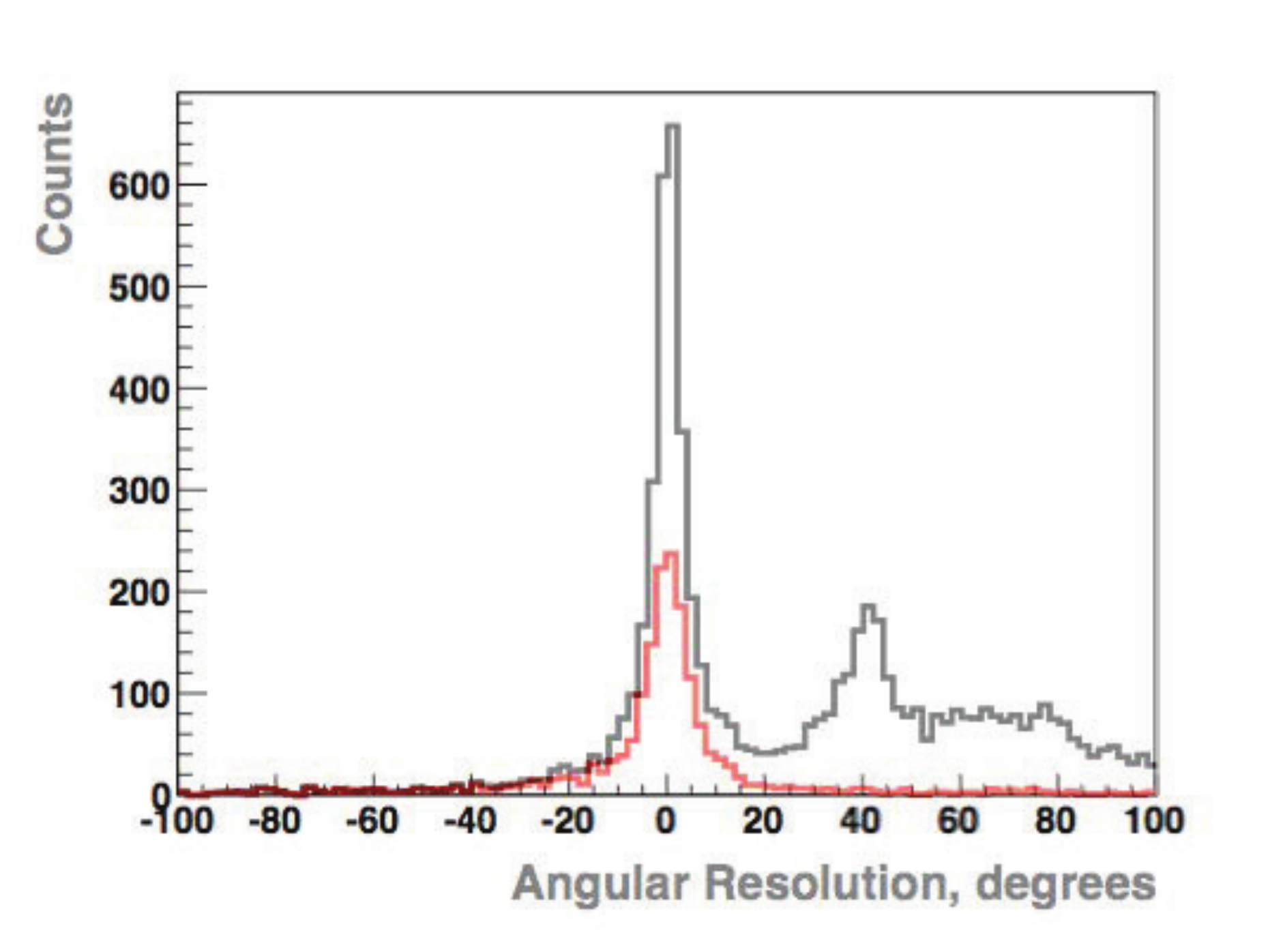}
\caption{Direction reconstruction for normal incidence 800 keV photons. Red line shows distribution for tracked events, and black line -  both tracked and untracked events (see explanation in the text).  }
\label{fig:figure2}
\end{minipage}
\end{figure}

 \subsection{Calorimeter}
In order to provide necessary information for both Compton and pair-production events over a broad energy range, the Calorimeter needs to have both good spatial and energy resolution over a broad energy range.
 For the parts closest to the Tracker, where knowing the positions of the interacting low-energy Compton-scattered photons is important, the Calorimeter should be position sensitive with $\sim$5mm position resolution in all 3 dimensions and have energy detection threshold less than 50 keV,
 In order to measure the higher-energy pair-production particles, a deeper Calorimeter is needed. Potentially it can be made less expensive with coarser position resolution (1-2 cm) with a higher energy detection threshold of  $>$100 keV. However, the thicker a precise coordinate-sensitive section of the calorimeter is (but more expensive), the better the instrument performance for Compton events (see Section 3 and Fig. 5, right panel).

The ComPair calorimeter is divided into 3 sections (Fig. 1). CZT-strip calorimeters at the bottom and sides detect low-energy Compton-scattered photons, while below the bottom-most CZT detector is a Fermi LAT-like CsI hodoscopic log calorimeter to provide energy measurement and reconstruction of longitudinal and latitudinal shower profiles for higher-energy pair-production events. The total thickness of the bottom section of the calorimeter
is $\sim$4.5 radiation lengths.
The CZT calorimeters are made of 4 planes of 5mm thick double-sided CZT-strip detectors. Each plane is made of 2cm x 2cm x 0.5cm individual CZT detectors, with four orthogonal readout strips on both sides. The strip pitch is 5mm, and strip width is 2.5mm. In the bottom section each plane is divided in 4 segments, similarly to the Tracker plane (Fig. 3). Each segment contains 25 x 25 individual daisy-chained CZT detectors, read out from two adjacent sides (tentatively by IdEAS VATA). 

The CsI calorimeter consists of 5 planes (or trays) of 1.2cm x 1.2cm x 32cm CsI(Tl) logs, with each log read out from both ends by Silicon Photomultiplier (tentatively Hamamatsu S12572-010P). In the top-most plane (tray) the logs are arranged along the X-axis, in the next one - along the Y-axis, and so on alternatively. The physical location of each CsI(Tl) log provides two spatial coordinates for the "hit", or center of gravity of energy deposition in this log. The one is the number of the plane, or Z-coordinate, and the other is the Y-position of the log in the X-tray, or alternatively X-position for the the Y-tray. The third coordinate along the log length is obtained from the asymmetry in light collection from each end of the log, with expected accuracy of $<$5 mm. 
An important feature of the ComPair calorimeter is the ability to reconstruct the shower profile using measured deposited energy in each log and its coordinates, similarly to how it is done in Fermi-LAT \cite{glast}. This improves energy measurement at higher energies by providing the ability to correct for energy leaking through the bottom and sides of the calorimeter. The shower shape is also an important discriminant between the electromagnetic showers produced by gamma-rays and hadronic showers produced by cosmic-ray background. 

\subsection{Anti-Coincidence Detector and Trigger}
The entire upper part of the instrument is covered by a plastic scintillator ACD, similar to that successfully used in Fermi-LAT. The main distinctive feature of the Fermi-LAT ACD is that it is divided in 89 segments in order to reduce a "back-splash" effect at high energy (above 100 GeV). It will not be a problem for ComPair due to the lower energy of interest, so the ACD design is significantly easier.
The baseline trigger is a five-fold coincidence of the hits in two consecutive Si planes (both sides in each plane) and the presence of a signal from the calorimeter, either CZT or CsI section, with no veto signal in the ACD within the triggering time window of a few microseconds. This trigger configuration allows detection of both Compton and pair-production events. The requirement of hits in two consecutive Si planes assumes the detection of the scattered electron for Compton events. The trigger can be re-configured to require a hit in only one Si plane and in the calorimeter (three-fold coincidence), subject to the orbital event data rate. This trigger configuration allows for the detection of "untracked" Compton events (see next Section).

\begin{figure}
\hspace{1cm}
\begin{minipage}[t]{0.88\textwidth}
\centering
\includegraphics[width=\textwidth]{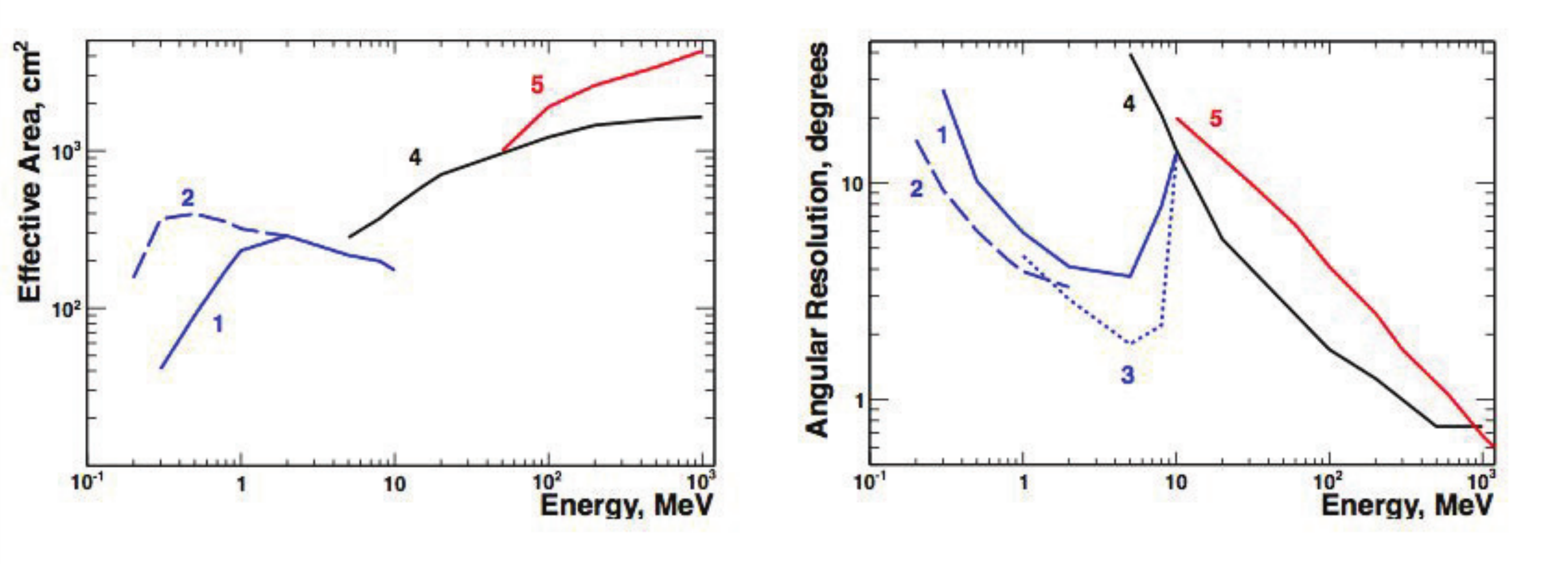}
 \caption{Left panel: Effective area. Right panel:  Angular resolution, FWHM for Compton events and 68\% containment radius for pair events. Line 1: Compton tracked events; line 2: Compton untracked events; line 3: Compton tracked events with deeper CZT calorimeter; line 4: pair events with ComPair; line 5: pair events with Fermi-LAT, front.}
\end{minipage}
\end{figure}

\section{Instrument Performance}
Instrument performance was simulated with MEGAlib - the Medium-Energy Gamma-ray Astronomy library \cite{megalib}  and cross-checked with MGEANT  \cite{mgeant}. MEGAlib is based on GEANT4 and has been successfully used for numerous simulations, including projects similar to ComPair: MEGA, GRIPS and ASTROGAM. It consists of two main logical parts: one simulates the events, and the other one reconstructs the events and analyzes the results. 
Event reconstruction for the pair events is rather straightforward, but it is more difficult  for the Compton events. As was stated in the instrument trigger description, we will mainly use Compton "tracked" events with a scattered electron detected in at least two Si planes. Using such events we will have smaller effective area and higher detection threshold than for "untracked" events without detected Compton electron, but the direction reconstruction is more reliable (Fig.4). The sample of "untracked" events has many events with misreconstructed directions (seen on the right from the central peak in this figure) which constitutes a significant source of background. Example of such background is atmospheric gamma-radiation arriving mainly from the horizon. However, for some specific observations, e.g. study of spatial structure of a source, the use of untracked events can be useful. Simulation results for effective area and angular resolution for normal incidence events, separately for Compton and pair events, are shown in Fig. 5, reflecting the effect of tracked and untracked events, along with characteristics for Fermi-LAT shown for comparison. We also investigated the effect of the thicker section of CZT calorimeter with better spatial resolution on the ComPair angular resolution. The dotted line (3) on the right panel of Fig. 5 shows angular resolution for the instrument with the CZT calorimeter section consisting of 10 planes, to be compared with the baseline design with 4 planes (solid blue line 1). The effect is visible, but the cost of such a calorimeter would be  significantly higher.

Fig. 6 shows the ComPair energy resolution. Clearly seen is the worsening of the resolution  at around 3-5 MeV  due to the events with a 511 keV annihilation photon escaping from the calorimeter. We have not implemented yet any shower shape reconstruction to improve energy reconstruction, so this is the conservative estimate of ComPair ability to measure the photon energy. The summary of ComPair expected performance and some technical parameters is shown in Table 1. \

\begin{figure}[ht]
\hspace{1cm}
\begin{minipage}[t]{0.45\textwidth}
\centering
\includegraphics[width=0.9\textwidth,trim=0 0 30 0 ,clip=true]{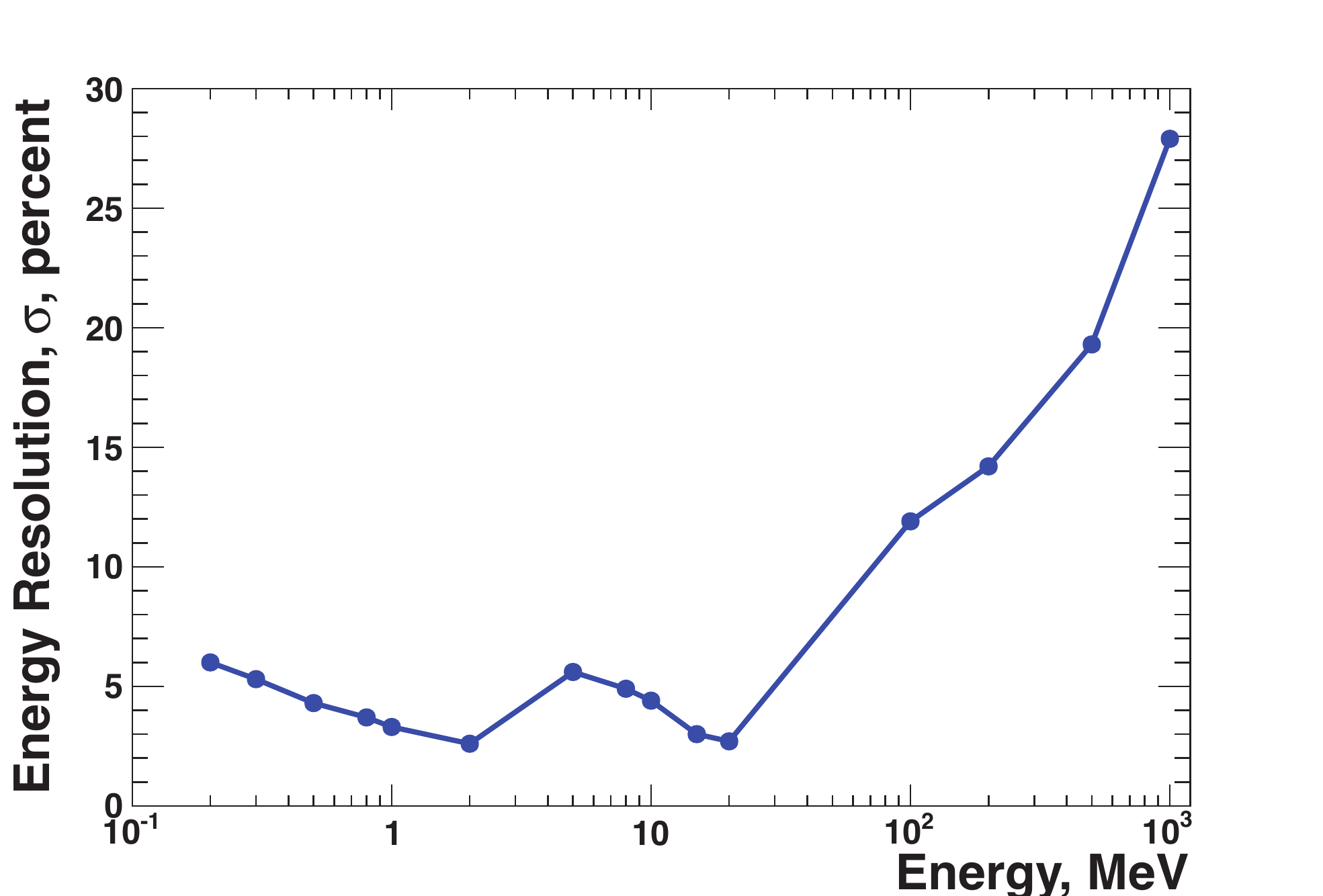}
\caption{Energy Resolution for ComPair}
\label{fig:figure1}
\end{minipage}
\begin{minipage}[t]{0.45\textwidth}
\centering
\includegraphics[width=0.9\textwidth,trim=0 0 30 0 ,clip=true]{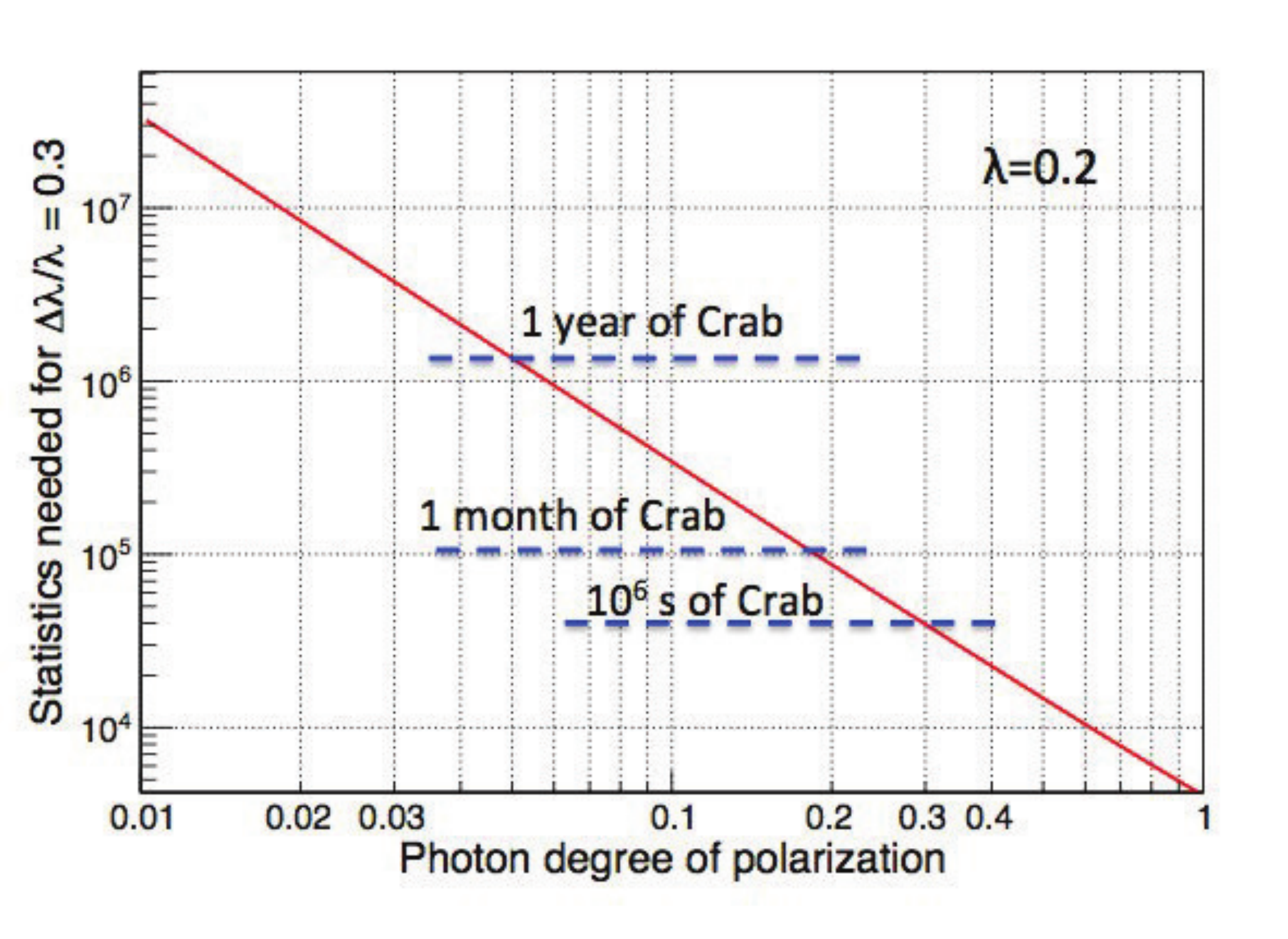}
\caption{The sensitivity of ComPair mission to detect polarization from the Crab (see explanation in the text)}
\label{fig:figure2}
\end{minipage}
\end{figure}

The source sensitivity  for the ComPair instrument compared with existing results is shown in Fig.8 (data for this figure are taken from  \cite{grips}  with added ComPair sensitivity). For Compton events we used only "tracked" events. For the energy below 30 MeV we expect an improvement in sensitivity by a factor of 20-50 compared with that of COMPTEL, and by a factor of 5-10 over EGRET in the energy range 30-100 MeV. 


In this paper we estimated the ComPair source sensitivity, assuming only isotropic Extra-Galactic Background, or EGB (\cite{gruber}, \cite{weiden}, \cite{fermi}).
However, the "natural" background such as  EGB for high latitude, and Galactic diffuse radiation for low latitude observations is not the main limiting factor in the instrument sensitivity. Compton events are also affected by the atmospheric and environmental (created in the instrument and spacecraft material) backgrounds. Accurate study of these, and possibly other sources of background, is the subject of a future work.  However we  estimated the systematic uncertainty, assuming five times higher background, to account for that sources. It is shown in Fig. 8 as a shadowed area above the ComPair line.  This is a preliminary estimate of the ComPair sensitivity, subject to future revision by adding more fidelity to the simulations and adding untracked events. 

\begin{figure}
\hspace{1cm}
\centering
\includegraphics[width=0.6\textwidth]{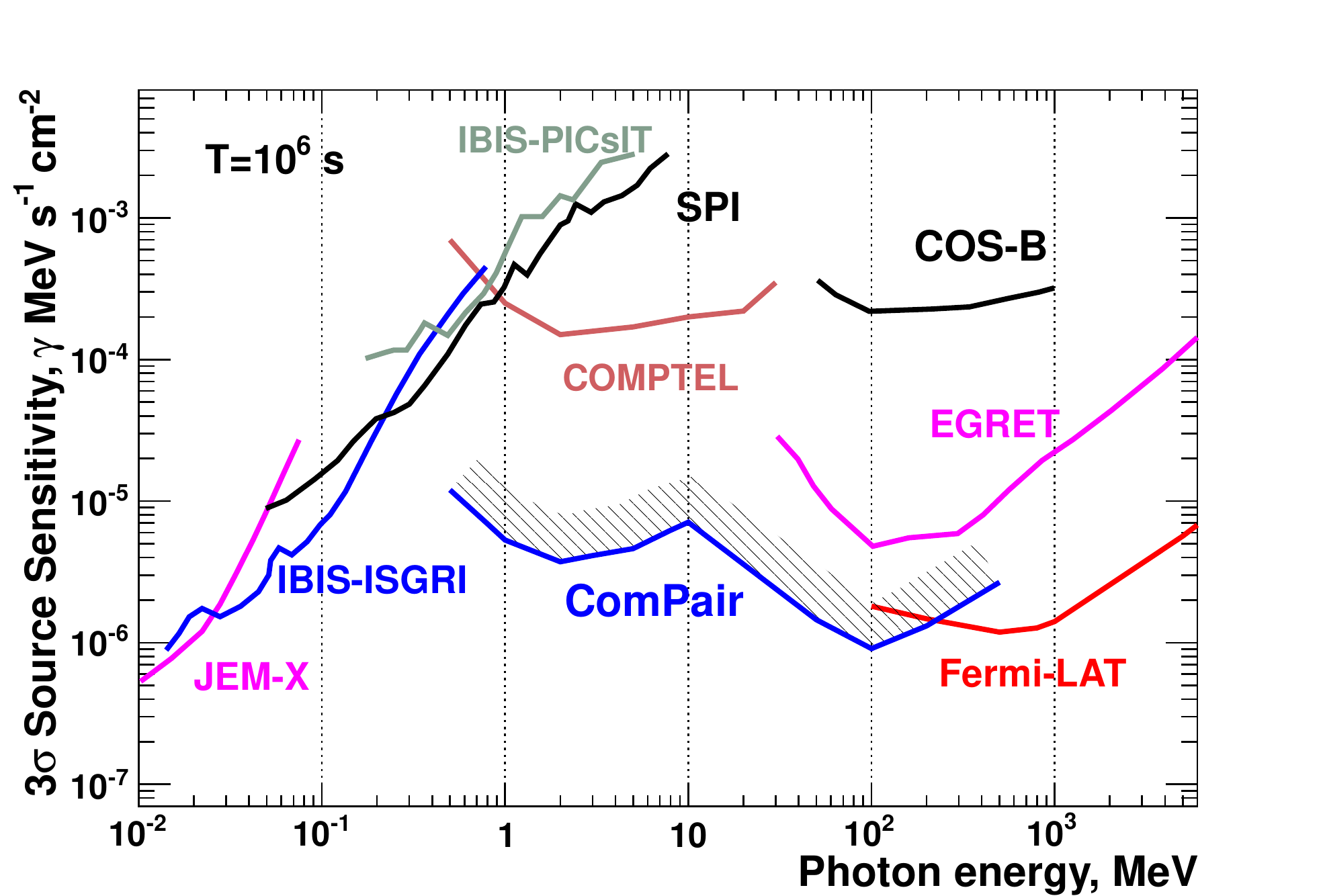}
\caption[width=\textwidth]{Point source sensitivity of previous gamma-ray measurements. The simulated sensitivity of ComPair mission is shown by the blue line, and shaded area shows systematic uncertainty}
\end{figure}

  \begin{table}[ht]
\centering
     \begin{tabular}{|c|c|}
\hline
   Energy Range & 0.5 - 100 MeV ( 0.2 - 500 MeV) \\
Energy Resolution & 2 - 5\% below 20 MeV, $\sim$ 12\% at 100 MeV \\
Effective Area & 50 - 250 $cm^2$ below 10 MeV, 200-1200 $cm^2$ above 10 MeV \\
Field of View & $\sim$ 3 sr \\
Angular Resolution & $\sim \; 6^{\circ} $ (1 MeV), $\sim \; 10^{\circ} $ ( 10 MeV), $\sim \; 1.5^{\circ} $ (100 MeV)  \\
Mass &  $< 1000$ kg \\
Power & $< 1000$ W \\
Overall dimensions & Diameter $\sim$ 200 cm, height $\sim$ 120 cm \\
    \hline
    \end{tabular}
     \caption{Summary of ComPair parameters}
     \label{tab1}
     \end{table}

The ComPair capability to detect radiation polarization from bright sources has been estimated (for pair events only) for the Crab pulsar based on calculations in \cite{kkl}. The major factor affecting the polarization measurement for pair events by pair-converter telescopes is the multiple scattering of the electron and positron in the converter, which is a Si-strip plane in ComPair. Although the Si-strip plane thickness of 0.5mm is not optimal, it still allows for such measurements. We estimated the ComPair sensitivity to measure linear polarization from the Crab, assuming measured asymmetry parameter $\lambda = 0.2$ (corresponding to the converter thickness). Red line in Fig. 7 shows how many events with energy greater than 10 MeV need to be detected to measure the source polarization with 30\% accuracy, and blue dash lines show the expected statistics to be collected by ComPair from the Crab for different observation time. For instance, if the Crab polarization is 20\%, it will be detected by ComPair in 1 month of observation time with 30\% accuracy. \
 
 \section{Summary}
We have developed a concept for a medium-energy gamma-ray space telescope, focused on accurate measurements in the energy range 500 keV - 100 MeV, with capability to perform measurements in an extended range from 200 keV to 500 MeV and higher. It inherits from predecessors such as Fermi-LAT, AGILE and MEGA. The ComPair concept assumes detecting capability for both Compton and pair production gamma-ray events and exceeds the currently achieved sensitivity of COMPTEL by a factor of $\sim$100 at energy around 1 MeV. The design presented in this paper can be considered as a concept with the possibility to scale it up or down depending on the available mission resources. 
The authors would like to thank Robert Johnson, Bill Atwood, Eric Grove, Eric Wulf, Igor Moskalenko and Steve Sturner for valuable discussions and suggestions. Also the authors are grateful to the MEGA team, and especially to Andreas Zoglauer for the wonderful simulation package MEGAlib and very valuable comments.

\end{document}